\documentclass[conference]{IEEEtran}
\IEEEoverridecommandlockouts

\usepackage{times}
\usepackage{graphicx}
\usepackage{cite,xspace,url,setspace}
\usepackage[cmex10]{amsmath}
\usepackage{amsthm}
\usepackage{amsmath}
\usepackage{algorithmic}
\usepackage{array}
\usepackage{bm}
\usepackage{hhline}

\usepackage[tight,footnotesize]{subfigure}
\usepackage{url,caption, color, enumerate}

\title{Real-Time and Embedded Deep Learning on FPGA for RF Signal Classification \thanks{This effort is partially supported by the U.S. Army Research Office under contract W911NF-17-C-0090. The content of the information does not necessarily reflect the position or the policy of the U.S. Government, and no official endorsement should be inferred.}
\thanks{\textsuperscript{\textcopyright} 2019 IEEE. Personal use of this material is permitted. Permission from IEEE must be obtained for all other uses, in any current or future media, including reprinting/republishing this material for advertising or promotional purposes, creating new collective works, for resale or redistribution to servers or lists, or reuse of any copyrighted component of this work in other works.}
\vspace{-3.7mm} 
}

\author{\IEEEauthorblockN{
Sohraab Soltani,
Yalin E. Sagduyu,
Raqibul Hasan,
Kemal Davaslioglu,
Hongmei Deng, and
Tugba Erpek}
\IEEEauthorblockA{Intelligent Automation, Inc., Rockville, MD, USA}
\IEEEauthorblockA{Email: \{ssoltani,  ysagduyu, rhasan, kdavaslioglu, hdeng, terpek\}@i-a-i.com}
\vspace{-7.5mm} 
}
\begin{document}
\newcommand{\argmax}{\arg\!\max}
\maketitle

\begin{abstract}
We designed and implemented a deep learning based RF signal classifier on the Field Programmable Gate Array (FPGA) of an embedded software-defined radio platform, DeepRadio\texttrademark, that classifies the signals received through the RF front end to different modulation types in real time and with low power. This classifier implementation successfully captures complex characteristics of wireless signals to serve critical applications in wireless security and communications systems such as identifying spoofing signals in signal authentication systems, detecting target emitters and jammers in electronic warfare (EW) applications,  discriminating primary and secondary users in cognitive radio networks, interference hunting, and adaptive modulation. Empowered by low-power and low-latency embedded computing, the deep neural network runs directly on the FPGA fabric of DeepRadio\texttrademark, while maintaining classifier accuracy close to the software performance. We evaluated the performance when another SDR (USRP) transmits signals with different modulation types at different power levels and DeepRadio\texttrademark receives the signals and classifies them in real time on its FPGA. A smartphone with a mobile app is connected to DeepRadio\texttrademark to initiate the experiment and visualize the classification results. With real radio transmissions over the air, we show that the classifier implemented on DeepRadio\texttrademark achieves high accuracy with low latency (microsecond per sample) and low energy consumption (microJoule per sample), and this performance is not matched by other embedded platforms such as embedded graphics processing unit (GPU).
\end{abstract}
\begin{IEEEkeywords}
Software-defined radio, deep learning, modulation classification, FPGA.
\end{IEEEkeywords}
\vspace{1mm} 
\section{Introduction}
The wireless communication environment is often characterized by high mobility, channel uncertainty, growth in traffic demands, and vulnerability to jamming. New agile technology solutions are much needed to characterize the spectrum and ensure reliable communications in such a fast-paced dynamic environment. Furthermore, spectrum resources are scarce across time, frequency, and space dimensions, and often shared among a variety of users and applications such as sensing, communications, and electronic warfare (EW). Cognitive radio has emerged as the enabling technology to make efficient use of spectrum resources and support adaptation of wireless communications in highly dynamic environments \cite{Haykin}. Supported by flexible software-defined radio (SDR) designs and implementations \cite{Magazine}, cognitive radio finds both commercial and tactical applications in conventional and emerging communications systems.  

Cognitive radio performs various tasks such as spectrum sensing and dynamic spectrum access (DSA) for situational awareness and spectrum agility.  To support these tasks, machine learning provides automated means to learn from and adapt to spectrum dynamics \cite{Chen2017, Simeone2018}. In particular, deep learning can process raw spectrum data and effectively operate on the latent representations by capturing and analyzing high-dimensional and dynamic spectrum data that conventional feature-based machine learning algorithms fail to grasp. 

Various waveform, channel, traffic, and interference effects, each with its own complex structures that quickly change over time, shape the spectrum \cite{Wang}. One critical step to assess the spectrum in terms of resources or vulnerabilities is to classify wireless signals. Signal classification is a key step of various tasks in wireless security and communications such as jamming/anti-jamming, device/RF fingerprinting, signal authentication, perimeter security, and interference hunting. Therefore, there is a growing interest in applying deep learning to signal classification such as modulation recognition \cite{Shea16}. Deep learning was applied to signals collected over the air \cite{Shea18}. However, the processing was confined to a host computer, which results in longer processing delay and limits it portability to embedded platforms. 

Wireless systems operate with high data rates, ranging from MHz in LTE to GHz in 5G systems. Therefore, high-volume data samples (in I/Q form) arrive at RF receivers to be processed. The latency to move the data to a host processor for deep learning is not feasible in real-time operations such as those that need to make a quick decision in microseconds time frame. While missing, a real-time embedded implementation of deep neural networks on the SDR is ultimately needed to support the embedded applications that run on stand-alone SDRs.
\begin{figure*}[t]
	\centering
	\includegraphics[width=1.5\columnwidth]{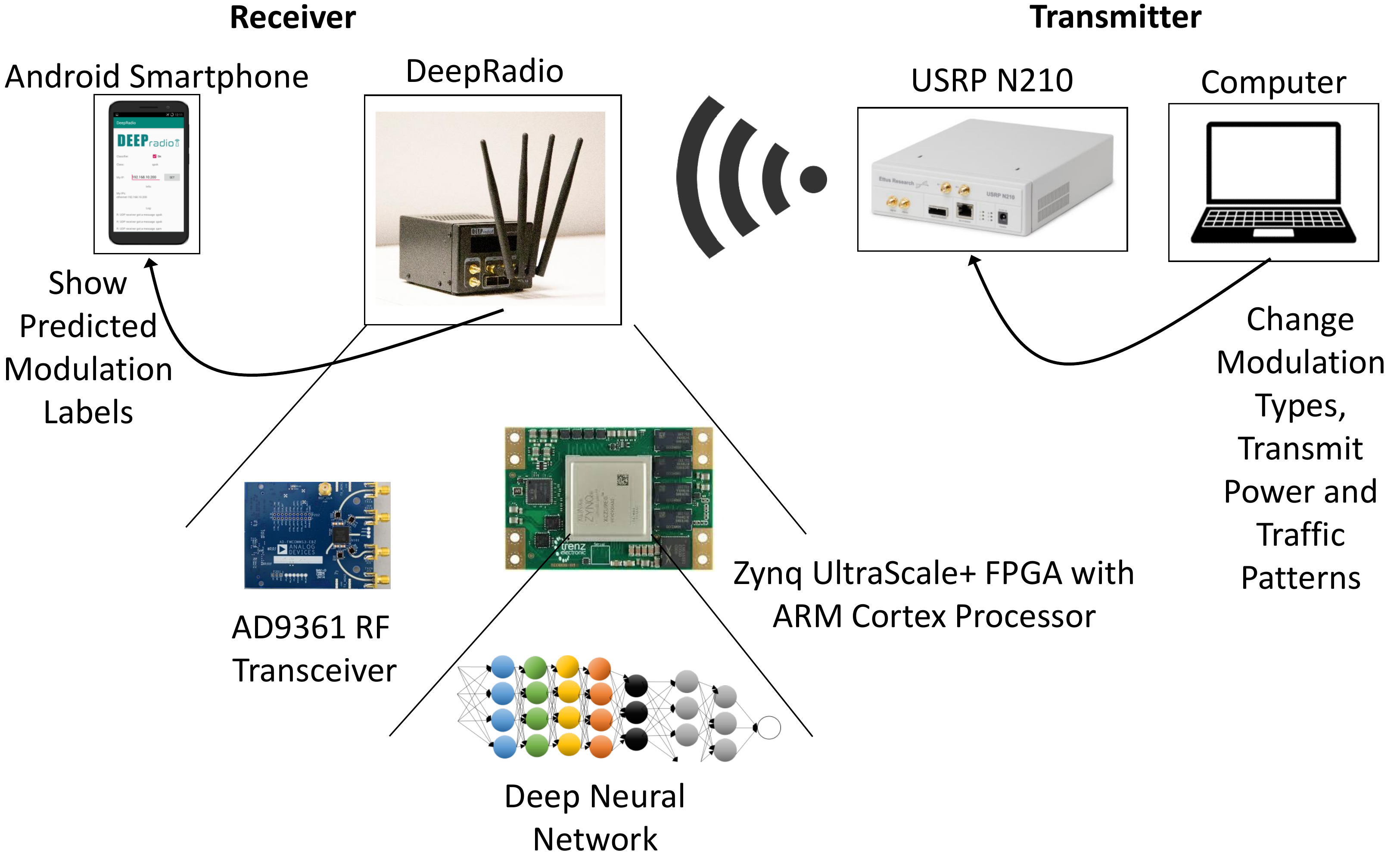}\\
	\caption{System setup for RF signal classification.}\label{fig:demosetup}
\end{figure*}

To fill this gap, we designed and implemented a deep neural network based classifier for signal classification on the  Field Programmable Gate Array (FPGA) of DeepRadio\texttrademark. Without using any other host processor or deep learning accelerator (such as \cite{accelarator}), the deep neural network runs directly on the FPGA fabric. This embedded solution provides high accuracy, low latency, and low power. In particular, we show the following novel capabilities for RF signal classification:
\begin{itemize}
	\item low-latency (microseconds), matching the speeds of RF spectrum dynamics in terms of channel, interference, and traffic,
	\item low energy consumption (microJoule), resulting in a longer battery lifetime (translated to longer network lifetime) and high portability, and
	\item high accuracy, approaching the limits of floating-point software operation.
\end{itemize}

Note that our goal is not to introduce another SDR. We use DeepRadio\texttrademark as an FPGA-based RF platform and provide a flexible design that can be ported to any other SDR that is equipped with FPGA. Our paper has three major discriminators compared to the state of the art that typically considers offline software implementation of simulated or prerecorded data:
\begin{itemize}
	\item We consider hardware generated and over the air transmitted physical data.
	\item We consider algorithm implementation on FPGA in an embedded hardware platform for low latency and low power consumption. 
	\item We consider real-time operation for both data collection and algorithm run.
\end{itemize}

The rest of the paper is organized as follows. Section II discusses related work. Section III introduces the system setup. Section IV presents implementation for deep neural network on FPGA. Section V presents the  results in terms of classification accuracy, latency, and power efficiency.

\section{Related Work}
Deep learning finds rich application in wireless communications. Examples include spectrum sensing \cite{Kemal2018}, MIMO detection \cite{HeMIMODet}, channel estimation and signal detection \cite{DeepOFDM}, physical layer communications \cite{OSheaTCCN}, jammer detection \cite{Yi2018}, stealth jamming \cite{Yi2018-2, IoT}, power control \cite{Terpek18}, signal spoofing \cite{Spoofing}, and transmitter-receiver scheduling \cite{Nof}.

RF signal classification can support different applications such as  radio fingerprinting \cite{finger} that can be ultimately used in cognitive radio systems \cite{sohraab} subject to dynamic and unknown interference and jamming effects \cite{jamming}. 
Modulation classification has been extensively studied  with deep neural networks \cite{Shea16,Shea18, Kim16,Mendis16,Peng17, Tu18, Rajendran2018, deVrieze18, Ali17, dyspan}, where the goal is to classify a given signal to a known modulation type. Different types of datasets have been used to train deep neural network for modulation classification. For example, \cite{Shea16} provided a training dataset collected from GNU Radio without any real channel or hardware impairments. On the other hand, \cite{Shea18} provided a training dataset collected from over-the-air measurements of USRP radio transmissions. However, those studies have performed modulation classification offline in software environments. Our goal is to run modulation classification in an embedded platform in real time while accounting for latency and power efficiency requirements. 

\section{System Setup}
The system setup is shown in Fig.~\ref{fig:demosetup}. There are two major components, a transmitter and a receiver. 

\begin{itemize}
\item There is one USRP N210 SDR that is controlled by a computer as an RF front end. This USRP either waits or transmits narrowband signals at 2.4 GHz frequency.  Different transmit powers will be employed to generate different signal-to-noise-ratio (SNR) effects. The transmissions can be made over the air or over cables depending on spectrum management restrictions. Each transmitted signal is modulated with one of six different modulation types, namely 
\begin{enumerate}
\item Binary Phase Shift Keying (BPSK), 
\item Quadrature Phase Shift Keying (QPSK), 
\item Continuous Phase Modulation (CPM), 
\item Gaussian Frequency Shift Keying (GFSK), 
\item Quadrature Amplitude Modulation (QAM) 16, and 
\item Gaussian Minimum Shift Keying (GMSK). 
\end{enumerate}
\item As receiver and classifier, DeepRadio\texttrademark runs a signal classifier by taking the received signals (I/Q samples) as input data and determines whether the received signal is noise (label 0) or it is a signal transmitted with one of the six modulation types (labels 1-6). 
\end{itemize}

Note that the uniqueness of the system set up is at the receiver side that makes real-time decisions by running a deep learning classifier on the FPGA of DeepRadio\texttrademark.

\begin{figure} [h]
	\centering
	\includegraphics[width=0.8\columnwidth]{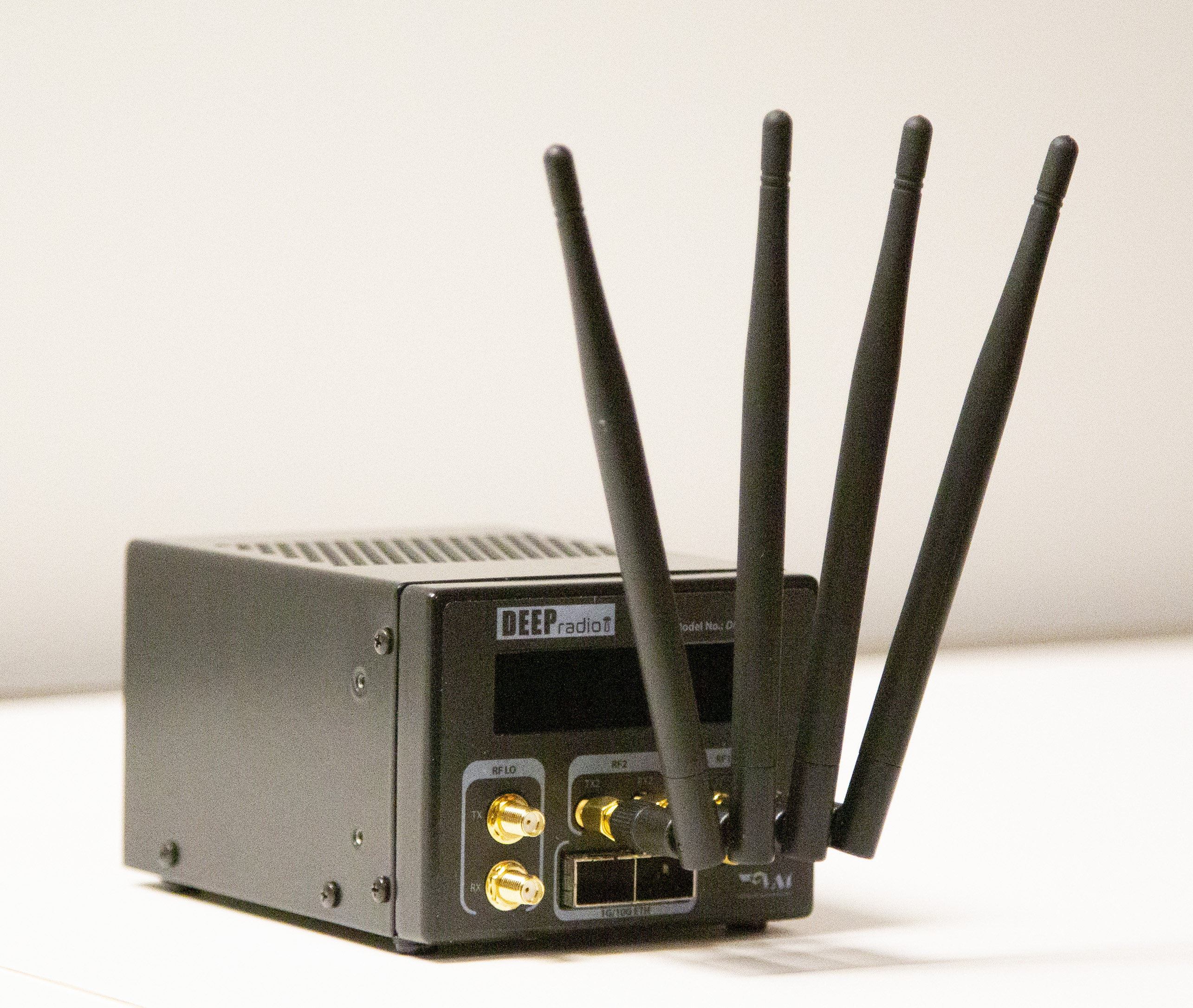}
	\caption{DeepRadio\texttrademark used for the RF classifier implementation.}\label{fig:radio}
\end{figure}

The received signals are passed to the FPGA and run through a deep neural network that returns the signal labels (noise or one of six modulation types). DeepRadio\texttrademark shown in Figure \ref{fig:radio} is equipped with Analog Devices AD9361 RF transceiver and Xilinx Zynq UltraScale+ XCZU9EG FPGA with ARM Cortex Processor (Quad-core ARM Cortex-A53 1.2 GHz, 64 bit). The FPGA has $600$ K System Logic Cells, $32.1$ Mb Memory, $2,520$ DSP Slices, and $328$ Maximum I/O Slices. The MIMO-capable RF transceiver covers $70$ MHz to $6$ GHz with tunable instantaneous bandwidth of between $200$ kHz to $56$ MHz.

Performance with the FPGA is compared to an embedded graphics processing unit (GPU) that connects to DeepRadio\texttrademark to receive I/Q data and runs the RF classifier. For that purpose, we use two types of embedded GPU from NVIDIA, namely Jetson AGX Xavier Developer Kit ($512$-core Volta GPU with Tensor Cores) and Jetson Nano Developer Kit ($128$-core Maxwell). 

 \begin{figure*} [t]
	\centering
	\includegraphics[width=1.3\columnwidth]{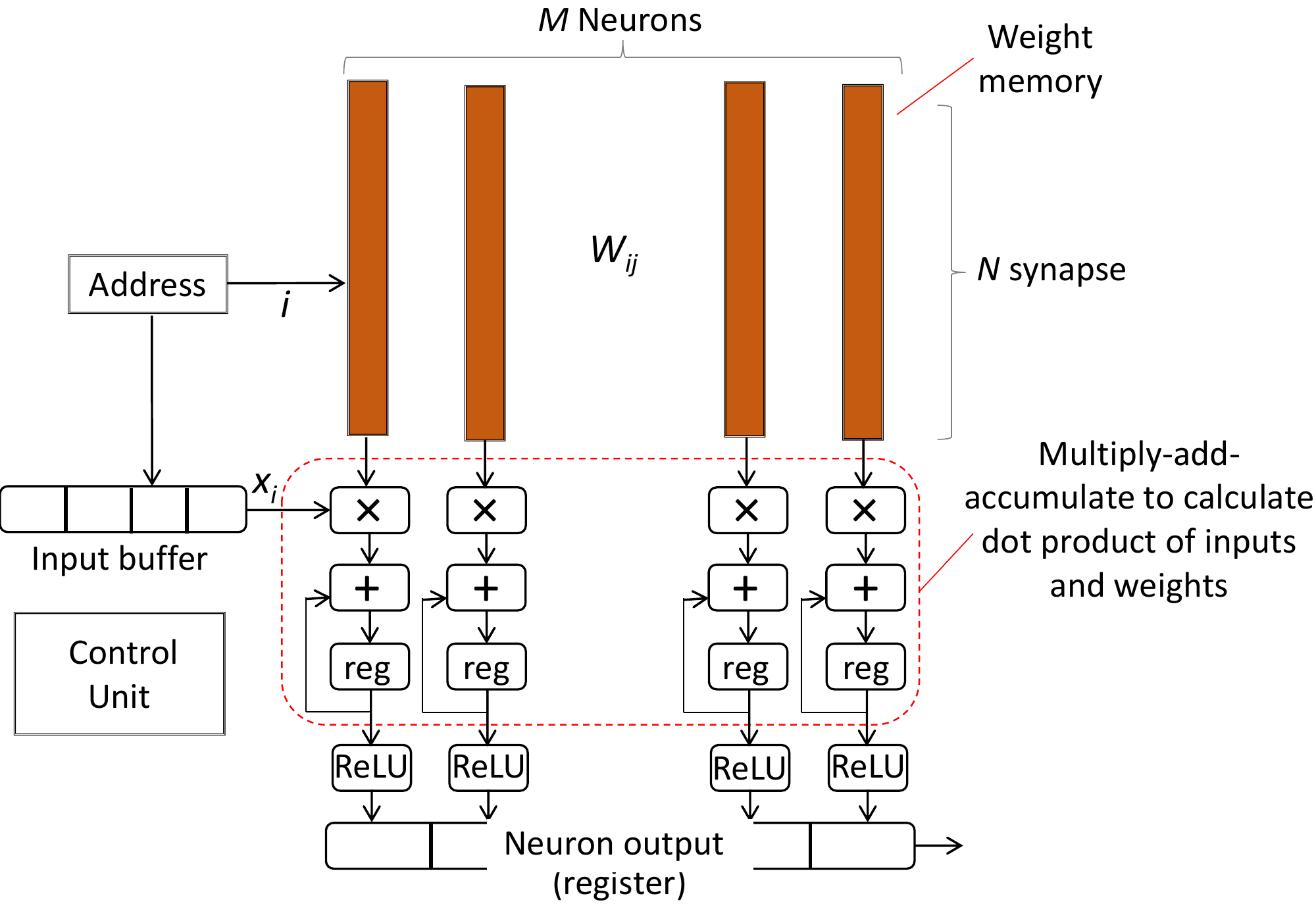}
	\caption{Block diagram of FNN implementation.}\label{fig:FNN}
\end{figure*}
\section{Implementation of Deep Learning based RF Signal Classifier}

The FPGA implementation focuses on inference (test) time. Hence, the training data was collected offline and a deep neural network was trained offline.  The trained model is a feedforward neural network (FNN). The input layer receives 900 I/Q signal samples that constitute one data sample for which a modulation label is assigned. 

The deep neural network is a function $F(\textbf{x})=\textbf{y}$ that takes an $n$-dimensional input $\textbf{x}\in \mathrm{R}^n$ (in our case $\textbf{x}$ is the received signal vector) and produces an $m$-dimensional output $\textbf{y} \in \mathrm{R}^m$ (in our case $\textbf{y}$ is the vector of likelihood scores corresponding to modulation types). For an $m$-class classifier, the output vector $\textbf{y}$ satisfies $y_1+\ldots+y_m=1$ and $1 \geq y_i \geq 0$. The classifier assigns a label $C(\textbf{x}) = \argmax_i F(\textbf{x})$. An activation function, denoted by $\sigma(\cdot)$ is applied at layer~$i$ for weights $\boldsymbol{\theta}_i$ and biases $b_i$ to perform $\sigma(\boldsymbol{\theta}_i \textbf{x} + b_i)$ operation. Let $F_i$ denote such operation at each layer, $\sigma(\boldsymbol{\theta}_i \textbf{x} + b_i)$, then a $k$-layer neural network can be represented as $F = \mathrm{softmax} \circ F_k \circ F_{k-1} \circ \ldots \circ F_1$ \cite{Carlini}. The neural network training tries to minimize a cost function $J(\boldsymbol{\theta})$ using the backpropagation algorithm by computing the gradient of the cost function with respect to neural network parameters, i.e., $\nabla_{\boldsymbol{\theta}} J(\boldsymbol{\theta})$. After hyperparameter optimization, 
the deep neural network architecture consists of four layers. The number of neurons is 1800 at the input layer, 100 and 20 at the first and second hidden layers, respectively, and 7 at the output layer. The  Rectified linear unit (ReLU) activation function is used at hidden layers. ReLU performs the $f(x) = \max(0,x)$ operation on input $x$. 
 Softmax activation function is used at the output layer. Softmax performs $f(\bm{x})_i = {e^{x_i}}/{\sum_j e^{x_j}}$ on input $\bm{x}$.
 The deep neural network is trained with the backpropagation algorithm \cite{Backprop} in TensorFlow \cite{Tensorflow} using crossentropy as the loss function. Cross-entropy function is given by 
 $\mathcal{L} = -\sum_{i=1}^m \beta_i \log(y_i)$, where $\boldsymbol{\beta} = \{ \beta_i \}_{i=1}^m$ is a binary indicator of ground truth such that only the index corresponding to correct label in $\boldsymbol{\beta}$ is $1$ and others are $0$. The predicted outputs by the neural network are denoted by $y_i$'s. Adam optimizer  \cite{AdamOpt} is used to update network weights iteratively based on training data.
 

 Figure  \ref{fig:FNN} shows the block diagram of the system implementing a layer of neurons in the FNN on the FPGA. Each layer of the FNN performs essentially same operations except with different numbers of neurons and synapses. Each neuron, in a FNN evaluates the dot-product of the inputs and its weights. As the neurons in a layer do not depend on each other for their operations, we perform the operations of each neuron in parallel. Parallel execution of the neurons provides low latency execution of the FNN on the FPGA. We implemented the ReLU activation on FPGA using conditional operation.

\begin{figure} [t]
	\centering
	\includegraphics[width=0.65\columnwidth]{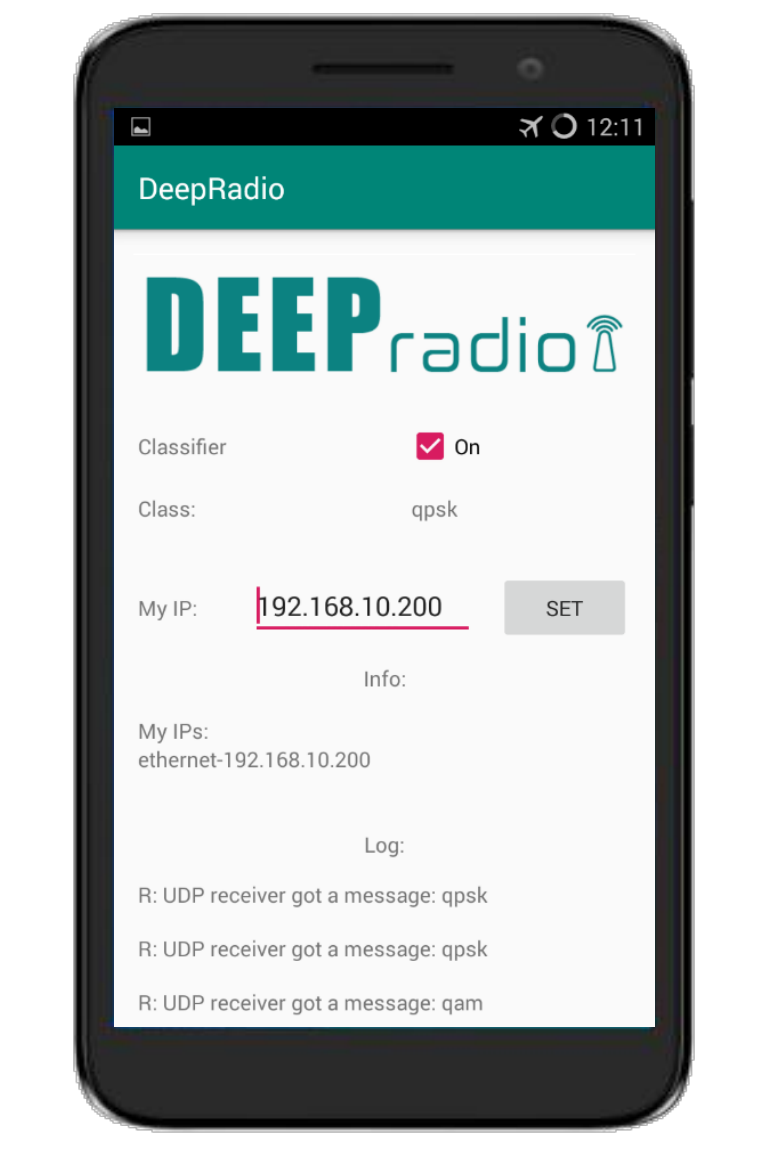}
	\caption{RF classifier GUI.}\label{fig:GUI}
\end{figure}

  The resulting TensorFlow model is converted to a format readable by FPGA and the bit file is generated. This bit file is uploaded to the FPGA. This way, the RF signal classifier runs on the FPGA without incurring any delay due to another host machine. The output label returned by the FPGA is passed to an Android smartphone for display. A mobile app is running on the smartphone to display the predicted label. The graphical user interface (GUI) of the smartphone is shown in Figure~\ref{fig:GUI}. 

\section{RF Classification Performance}

We measured that the RF signal classifier implemented on the FPGA achieves high accuracy  ($>94\%$ when averaged over different SNRs) with low latency and low energy consumption. Vivado Design Suite \cite{Vivado} is used to simulate and then synthesize the FPGA code. We port this code to FPGA and obtain classification results in real time from FPGA. 

\begin{figure} [h]
	\centering
	\includegraphics[width=0.95\columnwidth]{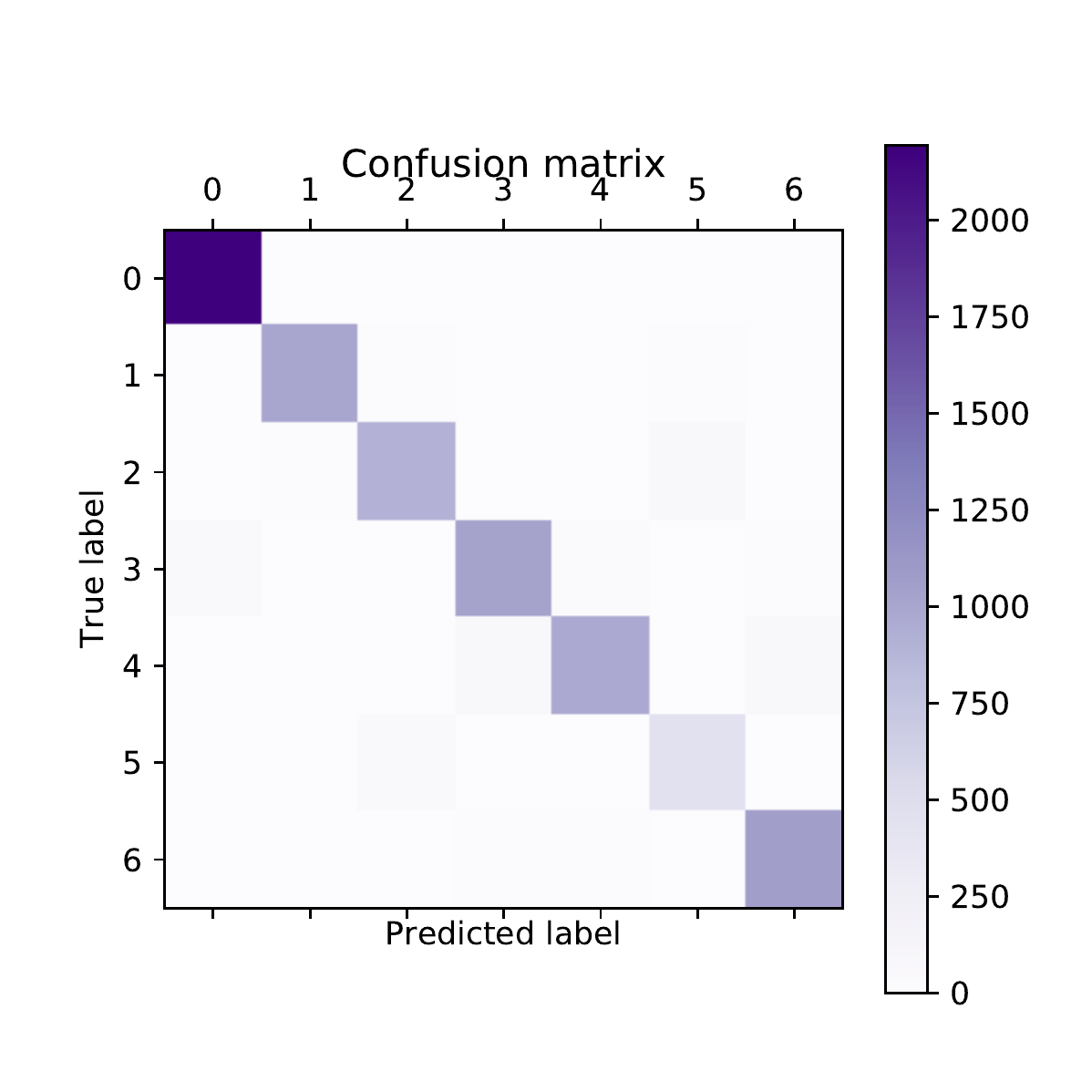}
	\caption{Confusion matrix of RF signal classifier.}\label{fig:confusion}
\end{figure}

The confusion matrix is shown in Figure~\ref{fig:confusion}, where label 0 is noise, label 1 is, BPSK, label 2 is QPSK, label 3 is CPM, label 4 is GFSK, label 5 is QAM16, and label 6 is GMSK. The average probability of error is shown in Table \ref{table:error} for each ground truth label. 

\begin{table*} 
	\caption{Average probability of error in classifying signals for each ground truth label.}
	\centering
	\setlength\doublerulesep{0.75pt}
	{\small
		\begin{tabular}{c|c|c|c|c|c|c|c}
			Ground Truth Label  & noise & BPSK & QPSK & CPM & GFSK & QAM16 & GMSK  \\ \hhline{=|=|=|=|=|=|=|=}
			Average Error of Misclassification &  $3.2\%$ & $8.2\%$ & $8.6\%$ & $13.9\%$ & $13.9\%$ & $0.10\%$ &  $3\%$
		\end{tabular}
	}
	\label{table:error}
\end{table*}


FPGA resource allocation is measured in Vivado. Resource allocation breakdown for RF signal classifier is shown in Table~\ref{table:resource}, where LUT: lookup table, LUTRAM: lookup table RAM FF: flip flop, DSP: digital signal processing, IO: input output, BUFG: global buffer.

\begin{table} 
	\caption{Average probability of error in classifying signals for each ground truth label.}
	\centering
	\setlength\doublerulesep{0.75pt}
	{\small
		\begin{tabular}{c|c|c|c}
			Resource  & Utilization & Available & Utilization (\%)  \\ \hhline{=|=|=|=}
			LUT & 158,435 & 274,080 & 57.81 \\ \hline  
			LUTRAM & 117,380 & 144,000 & 81.51 \\ \hline 
			FF & 16,222 & 548,160 & 2.96	\\ \hline 
			DSP & 210 & 2,520 & 8.33 \\ \hline 
			IO & 10 & 328 & 3.05 \\ \hline 
			BUFG & 4 & 404 & 0.99
		\end{tabular}
	}
	\label{table:resource}
\end{table}

\begin{figure*} [h]
	\centering
	\includegraphics[width=1.75\columnwidth]{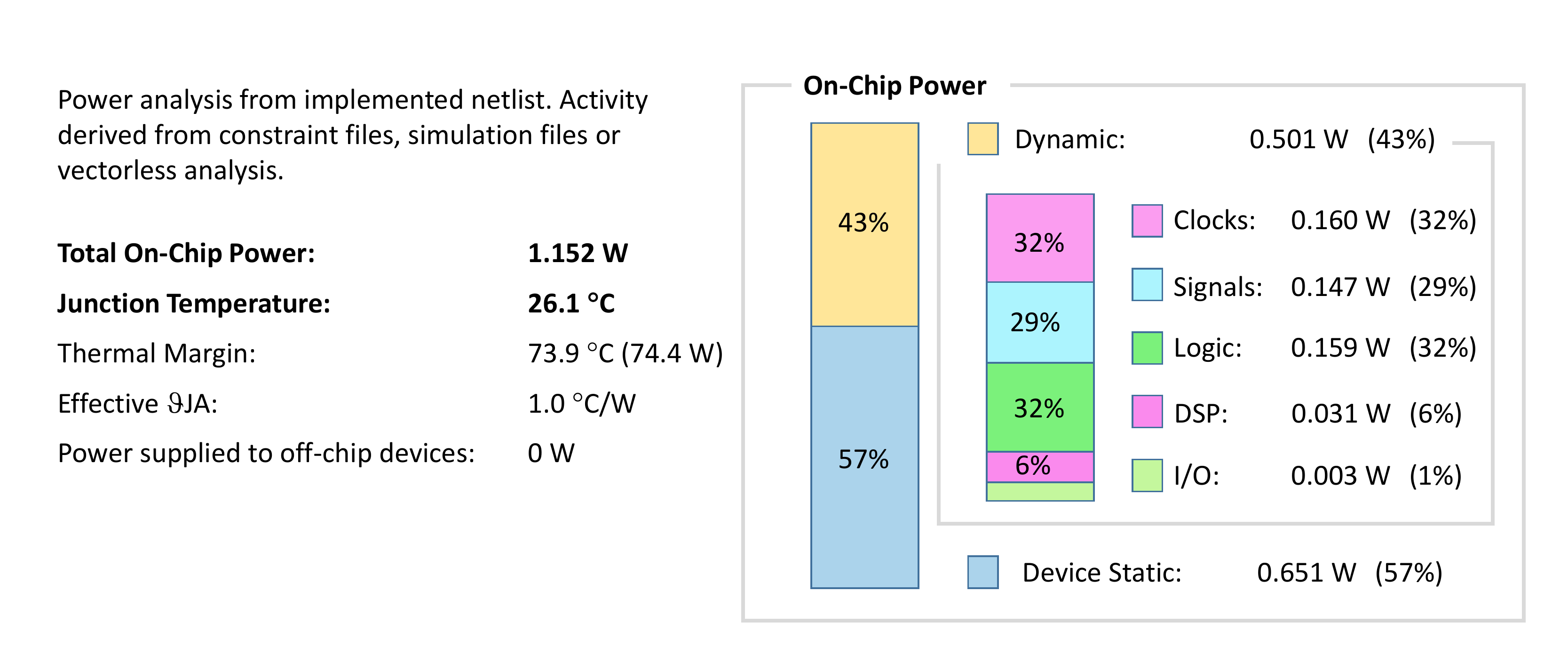}
	\caption{Power consumption on FPGA.}\label{fig:power}
\end{figure*}
We have the $16$-bit implementation of deep neural network on FPGA and the performance gap from the floating implementation in software (on embedded GPU) is within $1\%$. When the RF classifier is run on the FPGA, the latency is $24$ $\mu$s per sample and the energy consumption is $28$ $\mu$J per sample. Note that $24$ $\mu$s translates to operating with $>$$37$ megasamples per second without any downsampling. It is possible to increase the rate by downsampling, reducing the size of FNN, or using the available FPGA resources to implement another classifier in parallel on the FPGA.  Figure \ref{fig:power} shows the detailed breakdown of power consumption obtained from Vivado.

FPGA achieves better latency and power performance compared to other embedded platforms such as embedded GPU.
When the RF classifier is run on the Jetson AGX Xavier GPU, the latency is $3.6$ ms per sample and the energy consumption is $36$ mJ per sample. When the RF classifier is run on the Jetson Nano GPU, the latency is $4.1$ ms per sample and the energy consumption is $21$mJ per sample. Table \ref{table:GPU} shows the performance comparison of FPGA and embedded GPU.

\begin{table*} 
	\caption{Performance comparison of FPGA and embedded GPU.}
	\centering
	\setlength\doublerulesep{0.75pt}
	{\small
		\begin{tabular}{c|c|c|c}
			Performance Measure	$\backslash$ Device & FPGA & Xavier GPU & Nano GPU  \\ \hhline{=|=|=|=}
			Average Classification Accuracy & $94\%$ & $95\%$ & $95\%$ \\ \hline  
			Latency (per sample) & $24$ $\mu$s & $3.6$ ms & $4.1$ ms \\ \hline
			Energy Consumption (per sample) & $28$ $\mu$J & $36$ mJ & $21$ mJ
		\end{tabular}
	}
	\label{table:GPU}
\end{table*}

\begin{figure*} 
	\centering
	\includegraphics[width=1.9\columnwidth]{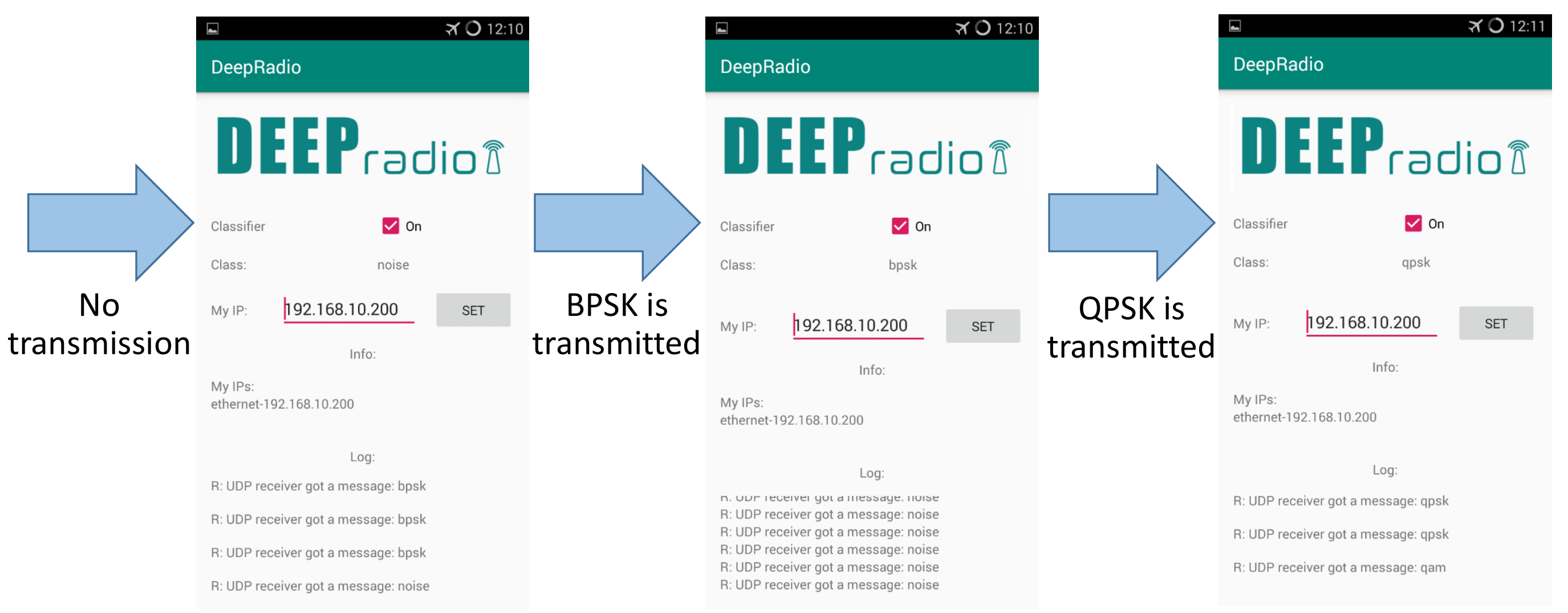}
	\caption{Recognized modulation is displayed in the smartphone GUI.}\label{fig:GUI_changes}
\end{figure*}

The predicted label is displayed in the smartphone GUI. Figure \ref{fig:GUI_changes} shows how the predicted label changes first when there is no transmission, second when a signal with BPSK is transmitted, and when third a signal with QPSK is transmitted.    

\section{Conclusion}
The paper presents the embedded implementation of deep learning based RF signal classification for low-latency and low-power applications. While deep learning has started finding different applications in wireless security and communications, there is a gap of real radio implementation. This paper fills the gap and opens up new opportunities regarding the use of deep learning in wireless applications. We showed that FPGA implementation of RF signal classifier maintains high accuracy and significantly reduces latency (more than $100$ times) and energy consumption (close to $1000$ times) compared to an embedded GPU implementation. With this capability, it is possible to support various wireless application such as detecting jammers/emitters, authenticating mobile devices, and  identifying spectrum opportunities all in real-time (in microsecond time frame) with low (microJoule) power consumption.    
  

\end{document}